\documentclass[amsmath,amssymb,12pt]{article}
\usepackage{jheppub}
\pdfoutput=1
\usepackage[utf8]{inputenc}
\usepackage{amsfonts}
\usepackage{graphicx}
\usepackage{slashed}
\usepackage{xcolor}

\title{Spectral weight and spatially modulated instabilities in holographic superfluids}
\author[1,2,3]{Blaise Gout\'eraux}
\affiliation[1]{Nordita, KTH Royal Institute of Technology and Stockholm University, Roslagstullsbacken 23, SE-106 91 Stockholm, Sweden}
\affiliation[2]{Stanford Institute for Theoretical Physics, Department of Physics, Stanford University, Stanford, CA 94305-4060, USA}
\affiliation[3]{APC, Universit\'e Paris 7, CNRS/IN2P3, CEA/IRFU, Obs. de Paris, Sorbonne Paris Cit\'e, B\^atiment Condorcet, F-75205, Paris Cedex 13, France (UMR du CNRS 7164).}
\author[2]{and Victoria L. Martin}
\emailAdd{blaise.gouteraux@su.se}
\emailAdd{vlmartin@stanford.edu}

\abstract{Free fermions form a Fermi surface, which results in non-zero spectral weight at low energy and finite wavevector $k_F$. In this work, we find similar features in holographic phases dual to strongly coupled quantum superfluid matter. At zero temperature, the phases we consider exhibit semi-local criticality in the IR and all the charge is carried by the scalar condensate outside the black hole horizon. Depending on the value taken by the IR critical exponents, we find Fermi surfaces in the transverse sector, Fermi shells in the longitudinal sector or no spectral weight at all. When there is non-zero transverse spectral weight, the IR can be subject to an instability at finite wavevector, the endpoint of which is likely a spatially modulated phase.}

\begin{document}
\subheader{NORDITA-2016-131, SU-ITP-16/23}
\maketitle
\section{Introduction}
Fermi liquid theory, a model explaining why a system of weakly-interacting fermions can exhibit properties similar to a non-interacting Fermi gas, is a mainspring in condensed matter physics. Weak coupling plays a crucial role, as the effective degrees of freedom are long-lived quasiparticles and can be thought of as dressed electrons. Although this theory describes the normal state of many metals, there also exist so-called non-Fermi liquids, characterized by an anomalous temperature scaling of the resistivity, specific heat, and other quantities while in their normal states \cite{varma1989phenomenology}. These materials are of significant technological interest, as they include the high-$T_c$ cuprates that become superconducting at unexpectedly high temperatures \cite{schofield1999non}. Many of these non-Fermi liquid features are believed to trace back to the lack of long-lived quasiparticles. It is now a major endeavor to move beyond Fermi liquid theory and develop an understanding of these strongly interacting materials. 

While standard perturbation theory techniques fail in the strong coupling regime, it is possible to investigate strongly interacting field theories using holography \cite{hartnoll2009lectures}. Despite their many scaling anomalies, non-Fermi liquids are seen to retain a Fermi surface momentum space distribution at low energies. Thus an important step in understanding these strongly correlated materials holographically is to first understand the contexts in which Fermi surfaces appear in holographic settings.
 
At strong coupling, the quantity that most explicitly marks the presence of a Fermi surface is called the low energy spectral weight:
\begin{equation}
\sigma(k)=\lim_{\omega\to 0} \frac{\text{Im}G^R_{JJ}(\omega,k)}{\omega}.
\end{equation}
It is denoted by $\sigma(k)$ to remind us that this is the low energy limit of the real part of the spatially-resolved conductivity appearing in Ohm's law (not to be confused with the optical conductivity where the $k\to0$ limit is taken first). The diagnostic power of the spectral weight is that it directly counts the number of degrees of freedom at a given frequency and momentum. By ``degrees of freedom" we mean the density of states into which a given state can transition. This counting property is easily seen from the spectral decomposition of the retarded Green’s function:
\begin{equation}
\text{Im}G^R_{JJ}(\omega,k)=\sum_{m,n}e^{-\beta E_m}|\langle n(k')|J(k)|m(k'')\rangle|^2\delta(\omega-E_m+E_n).
\end{equation}
In the expression above, $m$ and $n$ are eigenstates of the system. Notice that there are three momenta in the problem: $k$ (the perturbing momentum), $k'$ (the momentum of state $n$) and $k''$ (the momentum of state $m$). The counting comes from the explicit energy conserving delta function, and the momentum conserving delta function resulting from the inner product. Further, the appearance of the current operator $J$ guarantees that only charged degrees of freedom are counted. 

As we will discuss in the main body of the article, the low energy limit of the spectral weight is determined holographically by the IR behaviour of the holographic dual, that is by the near horizon, low temperature limit of the solution to the classical equations of the motion. Spectral weight in certain holographic theories has been studied previously, notably in gravity duals exhibiting a hyperscaling violating near horizon geometry (that is, geometry defined by the dynamical critical exponent $z$ and the hyperscaling violating exponent $\theta$) \cite{Charmousis:2010zz,Gouteraux:2011ce,huijse2012hidden}. 
These geometries describe a flow between a UV relativistic fixed point (with $z=1$) to an IR fixed point with emergent non-relativistic symmetry ($z\neq1$). Hyperscaling violation corresponds to an effective reduction of the spatial dimensionality of the IR fixed point.
For the hyperscaling violating geometries, it was found in \cite{hartnoll2012spectral} that spectral weight is always exponentially suppressed. The following year a particular limit of the hyperscaling violating geometry was studied, in which the quantity $\eta\equiv-\theta/z$ is held fixed as $z\rightarrow\infty$  \cite{anantua2012pauli}. The phase of matter dual to this theory is called the semi-local quantum liquid for its finite spatial correlation length but infinite correlation time \cite{Iqbal:2011in}. In both the hyperscaling violating and semi-local cases \cite{hartnoll2012spectral,anantua2012pauli}, the bulk matter content was given by the Einstein-Maxwell-dilaton theory \cite{Charmousis:2010zz,Gouteraux:2011ce}, and all electric flux is sourced by the charged black hole horizon (since there is no explicit charged matter in the bulk). According to the holographic dictionary, this electric flux is in turn the source for the field theory current density. This is the object of interest, allowing us to calculate the spectral weight. For the semi-local quantum liquid, nonzero low energy spectral weight was found to exist in the range $0<\eta<2$, for momenta less than a critical momentum:
\begin{equation}
\sigma(k)=\lim_{\omega\to 0}\frac{\text{Im}G^R_{JJ}(\omega,k)}{\omega}=\left\{
\begin{array}{ll}
\infty \qquad & k<k_{\star}\\ ~ 0\qquad & k>k_{\star}
\end{array}
\right.
\end{equation}
This is the signature of a (smeared out) Fermi surface. The infinite spectral weight at low momentum is an artifact of the zero temperature calculation. At nonzero temperature, this becomes finite (see \cite{anantua2012pauli} for further discussion). The dual charge density in the bulk gravity theory is behind the horizon, and we conclude that these charges are responsible for the low energy spectral weight observed holographically in the boundary field theory. 

In this work, we are interested in the low energy spectral weight of holographic superfluids. A holographic model of superconductivity was first established in \cite{gubser2008breaking, hartnoll2008building, hartnoll2008holographic}, and is reviewed in \cite{hartnoll2009lectures,hartnoll2011horizons}. The approach taken in \cite{hartnoll2008holographic} to study these theories was to consider an effective bulk action of the form 
\begin{equation}
S=\int d^4x \sqrt{-g}\left[R-2\Lambda-\frac14F^2-\left|\nabla\psi-i q A\psi\right|^2-V(|\psi|)\right].
\end{equation}
Here we have an explicit charged scalar in the bulk (outside the horizon) which can undergo spontaneous condensation below a critical temperature, and due to Schwinger pair production from the superconducting instability, all electric flux reaching the boundary is sourced by the charged condensate rather than the black hole horizon \cite{hartnoll2011horizons}. For a generic choice of a quadratic or quartic potential for the scalar potential $V$, the IR geometry is another copy of Anti de Sitter or a Lifshitz spacetime \cite{Gubser:2009cg,Horowitz:2009ij}. The general results of \cite{hartnoll2012spectral} indicate that in both cases, the low-energy spectral weight is exponentially suppressed as the exponent $z$ takes a finite value. That is, the low-energy spectral weight does not seem to depend on the opening of a superconducting gap at all, contrary to weak coupling intuition. Rather, what appears to govern its absence are the scaling properties of the IR spacetime. The main goal of this paper is to explore what happens in the cases where $z\to+\infty$, for which \cite{hartnoll2012spectral,anantua2012pauli} found non-zero low energy spectral weight. To be precise, weakly-coupled intuition dictates the spectral weight at finite momentum should vanish after the U(1) symmetry is spontaneously broken and the charged condensate has formed in the bulk, accompanying the expulsion of all the charge outside the horizon.

To explore this question, we will consider a slightly more general model \cite{gouteraux2012quantum,Gouteraux:2013oca}:
\begin{equation}
S=\int d^4x \sqrt{-g}\left[R-\frac12\partial\phi^2-\frac14Z(\phi)F^2-\frac12W(\phi)A^2-V(\phi)\right].
\end{equation}
This is an effective theory that captures the physics of superconductivity: the scalar has already undergone Bose-Einstein condensation, as evidenced by the massive photon. The scalar field $\phi$ should be thought of as the modulus of the original complex scalar field, while its phase has been integrated out.

This model allows for several interesting features, depending on the IR behaviour of the action couplings $V(\phi)$, $Z(\phi)$ and $W(\phi)$. First, the IR geometries now include semi-local fixed points with $z\to\infty$, even in the presence of a condensate. Second, these superfluid semi-local geometries can come with either charge in the bulk exclusively, or with both charge behind the horizon and in the bulk \cite{gouteraux2012quantum}. Here we will focus on the former case.

Our main result is that we do find nonzero low energy spectral weight at finite momentum in the boundary field theory, in spite of the fact that a condensate has formed and all the bulk charge now sits outside the horizon. The presence or absence of low energy spectral weight appears to depend on the scaling properties in the IR and not on the spontaneous breaking of the U(1) symmetry. This raises several interesting questions: 1) How should we interpret low energy spectral weight that exists independently of horizon charge? 2) What other degrees of freedom could this weight represent? 3) To what extent do bulk charge distribution properties represent those of the boundary charge?

In what follows, we will outline the calculation of the zero temperature low energy spectral weight for holographic superfluids in two decoupled sets of variables: the transverse and longitudinal channels. For the transverse channel we solve and decouple the bulk equations of motion directly, and for the longitudinal channel we demonstrate a scaling argument that allows us to infer the low-frequency dependence of the retarded Green's function without the need of a full bulk calculation. We end by a discussion of these results and suggest directions for future work. 

\section{Einstein-Maxwell-dilaton with massive vector}

In this work we consider the  Einstein-Maxwell-dilaton theory with a massive vector, which spontaneously breaks the $U(1)$ gauge invariance of the theory
\begin{equation}\label{EPDaction}
S=\int d^4x \sqrt{-g}\left[R-\frac12\partial\phi^2-\frac14Z(\phi)F^2-\frac12W(\phi)A^2-V(\phi)\right].
\end{equation}
The equations of motion of the theory in terms of a general Ansatz for the metric and matter fields are given in Appendix \ref{AppEoms}.

The low energy spectral weight is determined by the IR behaviour of the solution to the classical equations, so we focus only on solutions describing the near horizon region of spacetime at zero temperature.
They have the following runaway behavior deep inside the bulk
\begin{equation}
\label{runningscalar}
\phi\underset{IR}{\longrightarrow}\infty.
\end{equation}
This is a necessary condition to depart from the scale-invariant solutions considered in \cite{Gubser:2009cg,Horowitz:2009ij}.
It is known \cite{Iizuka:2012pn,gouteraux2012quantum,Gath:2012pg,Gouteraux:2013oca} that such a solution will emerge from the equations of motion of if we allow the coefficient functions to take the following form in the IR, loosely motivated by top-down string theory realization of the model \eqref{EPDaction}:\footnote{Top-down realizations of holographic superfluids with exponential scalar couplings can be found in \cite{Gubser:2009qm,Gauntlett:2009dn,Gauntlett:2009bh,Bobev:2011rv,DeWolfe:2015kma,dewolfe2016gapped}.}

\begin{equation}
V(\phi)=V_0 e^{-\delta\phi}\,,\qquad Z(\phi)=Z_0 e^{\gamma\phi}\,,\qquad W(\phi)=W_0 e^{\epsilon\phi}\,.
\end{equation}
For the rest of the analysis we are free to set $Z_0=1$, as this is simply a rescaling of the gauge field. 

More precisely, the scaling solutions will take the general form
\begin{equation}
\label{EMDHV}
ds^2=r^{\theta}\left(-\frac{dt^2}{r^{2z}}+\frac{L^2 dr^2+dx^2+dy^2}{r^2}\right),\quad A_t=Q r^{\zeta-z}\,,\quad \phi=\kappa\log(r)\,.
\end{equation}
The logarithmically running scalar manifestly realizes our condition \eqref{runningscalar}. The metric and gauge field are covariant under rigid rescalings of the coordinates $t\to\lambda^z t$, $(r,x,y)\to\lambda (r,x,y)$. Upon turning on a small temperature, the entropy density is seen to scale as $s\sim T^{(2-\theta)/z}$, which exemplifies how the spatial dimensionality of the boundary field theory has been effectively modified to $d_{eff}=2-\theta$ and characterizes hyperscaling violation in the IR field theory. In the bulk, the critical exponent $\zeta$ encodes departure from scale invariance for the gauge field. From the point of view of the dual theory, it governs the IR scaling of the electric conductivity \cite{Gouteraux:2013oca, Gouteraux:2014hca}. It also contributes anomalously to the dimension of the charge density operator in the IR \cite{Gouteraux:2013oca, Gouteraux:2014hca,Karch:2014mba} (though one should note that when the current is not conserved, its dimension is not protected and the usual arguments preventing an anomalous dimension \cite{PhysRevB.46.2655} do not apply).

The parameters of the solution \eqref{EMDHV} are fixed in terms of parameters in the action, so $(\theta,z,\zeta,L,\kappa)$ are actually functions of $(\epsilon,\gamma,\delta,W_0,V_0)$. In this paper, we will not study the general class of scaling solutions  \eqref{EMDHV} (for details see \cite{gouteraux2012quantum,Gouteraux:2013oca}) but instead focus on their semi-local limit $z\to+\infty$, $\theta\to+\infty$, $\zeta\to+\infty$ with $\eta=-\theta/z$ and $\tilde\zeta=\zeta/z$ kept finite. In the remainder of the paper, we drop the tilde.

Two cases may be distinguished, depending on whether the bulk charge sits entirely outside the horizon or also behind it. In this work, we will focus on the first possibility and refer to \cite{gouteraux2012quantum} for more details on the second.

The $\eta$ geometries are exact solutions to the background Einstein equations provided the constraint $\epsilon=\gamma-\delta$ is enforced \cite{gouteraux2012quantum,Gouteraux:2013oca}:
\begin{equation}\label{metric}
ds^2=r^{-\eta}\left(\frac{-dt^2+ dr^2}{r^2}+dx^2+dy^2\right),\qquad \phi(r)=\kappa\log r
\end{equation}
with
\begin{equation}\label{background param}
\begin{split}
&A=\sqrt{\frac{2}{1-\zeta}}r^{\zeta-1}\,,\quad V_0=\zeta-1-\eta-\eta^2, \quad W_0=(\zeta+\eta)(1-\zeta)\\
& \epsilon=\gamma-\delta\,,\quad \kappa\delta=-\eta\,,\quad\kappa\gamma=-(\eta+2\zeta)\,,\quad \kappa=\sqrt{\eta^2-2\zeta}\,.
\end{split}
\end{equation}
Note that these solutions do not exist for all values of $\gamma$, $\delta$ and $W_0$. Instead, choosing two fixes the third. This places constraints on the effective actions which admit these solutions. We have also chosen to set the IR `AdS radius' to $1$ by fixing $V_0$. This is of course not necessary (and actually not desirable to construct the full flow to a UV AdS$_4$).

It is important that we only consider a physically-consistent parameter space. The constraints we have to take into account are: the null energy condition (NEC), the reality of all metric/scalar/gauge field coefficients, negativity of $V_0$ and positivity of $W_0$ (since this is a mass term and the charge squared of the scalar \cite{gouteraux2012quantum}). We also require that the specific heat is positive. According to our convention, $V_0<0$ because the $V(\Phi)$ term in the action replaces the cosmological constant term $-2\Lambda$, and $\Lambda<0$ for Anti-de Sitter space. Furthermore, from the metric (\ref{metric}) we see that in order to have a well-defined IR we must have $\eta(\eta+2)>0$. This constraint is echoed in the NEC, which requires that the gravitational field generated by the stress tensor is attractive. Specifically, the NEC gives \cite{dong2012aspects}:
\begin{align}
\eta(2+\eta)>0 \\ \eta+1>0.
\end{align}
Combining all of our constraints, we find
\begin{equation}
\eta>0\,,\qquad -\eta<\zeta<\text{min}(\eta^2/2,1).
\end{equation}
From these conditions, it follows that the IR is always $r\to+\infty$ in these coordinates and the electric flux, which scales as $r^{-\zeta-\eta}$, always vanishes there.

Finally, we should also be careful that there exists the right number of irrelevant deformations around the IR geometry. These correspond to the scaling dimensions of operators. The method to compute the radial deformations and the identification of the IR scaling dimensions is described in detail in \cite{gouteraux2012quantum,Gouteraux:2013oca}. The deformations all sum to $1+\eta$, which is the dimension of the free energy density in the IR. There are three pairs of modes. Two of them are simply $(1+\eta,0)$. The other pair is the one of interest and should be used to connect the IR geometry to the UV AdS geometry. It reads
\begin{equation}
\beta^\pm=\frac12\left(1+\eta\pm\sqrt{\frac{-32 \zeta ^3+\zeta ^2 (48-32 \eta )-2 \zeta  \left(13 \eta ^2-10 \eta +9\right)+(\eta -3)^2 \eta ^2}{\eta^2-2\zeta}}\right) 
\end{equation}
We can see that $\beta^+>0$, so this is always a relevant mode. We need to impose that $\beta^-$ is irrelevant, which leads to the reduced parameter space
\begin{equation}
\label{ParSpace}
\left(0<\eta \leq \frac{1}{2} \left(\sqrt{5}-1\right)\textrm{and} -\eta <\zeta <\frac{\eta ^2}{2}\right)\textrm{or} \left(\eta >\frac{1}{2} \left(\sqrt{5}-1\right)\textrm{and} -\eta <\zeta <\frac{1-\eta }{2}\right)
\end{equation}
This is shown in Figure \ref{fig:ParSpaceTransverse}.

These solutions are called cohesive as the electric flux they source vanishes in the IR: the horizon is neutral and all the charge is generated by the charged condensate in the bulk.

\section{Computation of the spectral weight}

We now perturb the background fields 
\begin{equation}
g_{\mu \nu}\rightarrow g_{\mu \nu}+\delta g_{\mu \nu}, \qquad A_y\rightarrow A_y +\delta A_y
\end{equation}
 and determine the linearized equations of motion. If we choose the plane wave perturbations to be in the $x$-direction, as $\delta X(r,x,t)=\delta X(r)e^{i(kx-\omega t)}$, then the perturbed modes naturally decouple into two categories: those even or odd under the parity transformation $y\rightarrow -y$.  The set of odd modes are called transverse $\{ \delta A_y,\delta g_{xy},\delta g_{yt},\delta g_{yr}\}$ and the even modes are called longitudinal $\{ \delta A_t,\delta A_x,\delta A_r,\delta g_{tt},\delta g_{rt},\delta g_{xt},\delta g_{xr},\delta g_{xx},\delta g_{yy},\delta g_{rr},\delta \Phi\}$.

Not all of the perturbations mentioned above are independent, and we must combine them to form gauge invariant variables. The gauge group here is just diffeomorphism invariance, and so gauge invariant variables are those that are invariant under a coordinate transformation $X_{\alpha}\rightarrow X_{\alpha}+\xi_{\alpha}$. Under such a transformation, the metric changes according to the Lie derivative
\begin{equation}
\delta g_{\mu \nu}\rightarrow \delta g_{\mu \nu}+\nabla_{(\mu}\xi_{\nu)}. 
\end{equation}
Thus for a gauge invariant variable, the Lie derivative must vanish. Note that the choice of variables is not unique, though some choices are wiser than others as a tool for decoupling the equations of motion. Since there are four degrees of freedom in $\xi_\alpha$, we can pick a gauge such that four metric perturbations are zero. A popular choice is the so-called radial gauge $\delta g_{\mu r}=0$. However, it will turn out that postponing this gauge choice will allow us to more easily decouple the perturbed equations of motion in the transverse channel. 

The vector potential also transforms according to the Lie derivative:
\begin{equation}
\delta A_y\rightarrow\delta A_y+(\xi^\lambda\partial_\lambda \delta A_{y}+ \delta A_{\lambda}\partial_y\xi^\lambda).
\end{equation}
The vector $\xi$ is contracted, so this transformation gives us the freedom to set only one of the variables $\delta A_{\mu}$ equal to zero. A further discussion of Lie derivatives and gauge invariant variables is given in the Appendix.

\subsection{Transverse channel} \label{trans sec}
In the transverse channel the modes are:
\begin{equation}
\begin{split}
&\delta g_t^y=\delta g_t^y(r) e^{-i\omega t+i k x},\quad\delta g_r^y=\delta g_r^y(r) e^{-i\omega t+i k x},\\ 
&\delta g_x^y=\delta g_x^y(r) e^{-i\omega t+i k x},\quad\delta A_y=\delta A_y(r) e^{-i\omega t+i k x}.
\end{split}
\end{equation}
Note that it is sometimes more useful to work with perturbed variables with one index raised.

The resulting equations of motion (prior to substituting in the background quantities) are
\begin{equation}
0=k^2 \delta g_t^y+k\omega \delta g_x^y-A W \delta A_y-\frac{Z}{B}A'\delta A_y'-\frac1C\sqrt{\frac{D}{B}}\left[\frac{C^2}{\sqrt{BD}}\left(\delta g_t^y{}'+i\omega \delta g_r^y\right)\right]'
\end{equation}

\begin{equation}
0=ik(\delta g_x^y{}'-ik\delta g_r^y)+\frac{i\omega C}{D}(\delta g_t^y{}'+i\omega \delta g_r^y)+i\omega\frac{Z}{D}A'\delta A_y
\end{equation}

\begin{equation}
0=k\omega\frac{C}{D} \delta g_t^y+\omega^2 \frac{C}{D}\delta g_x^y+\frac1{\sqrt{BD}}\left[\frac{C\sqrt{D}}{\sqrt{B}}\left(\delta g_x^y{}'-ik \delta g_r^y\right)\right]'
\end{equation}

\begin{equation}
0=\frac1{\sqrt{BD}}\left[\frac{\sqrt{D}}{\sqrt{B}}Z \delta A_y'\right]'+\frac{CZA'}{BD}\left(\delta g_t^y{}'+i\omega \delta g_r^y\right)+\delta A_y\left(\omega^2\frac{Z}{D}-k^2\frac{Z}C-W\right).
\end{equation}
The perturbations we have introduced are not gauge invariant variables, and so our first task is to find linear combinations of the fields $\{ \delta A_y,\delta g_{x}^y,\delta g_{t}^y,\delta g_{r}^y\}\ $ that are gauge invariant. We introduce the vector field
\begin{equation}
\xi=e^{-i(kx-\omega t)}\xi^y(r)\partial_y
\end{equation}
and calculate the Lie derivative with respect to this field. We have chosen $\xi$ so that it has one nonzero component, the y-component. This is because we are working in the transverse channel, where the modes are odd under the parity transformation $y\rightarrow -y$. An explicit calculation is given in the appendix. The answers are:
\begin{align}
\begin{split}
&[\mathcal{L}_\xi g]^{y}_t=-i\omega \xi^y,\quad[\mathcal{L}_\xi g]^{y}_r=\xi^{'y}, \\
&[\mathcal{L}_\xi g]^y_{x}=ik \xi^y,\quad~~~[\mathcal{L}_\xi a]_{y}=0.
\end{split}
\end{align}
In general, beyond finding vanishing combinations of these quantities, one must also take care that no longitudinal modes $\{\ \delta A_t,\delta g_{rt},\delta g_{yy}, ...\}\ $ are generated in this process. In this case, that issue does not arise. We choose the combinations: 
\begin{equation}
\psi_1=r^{2-\eta/2}(\delta g_t^y{}'+i\omega \delta g_r^y)\,,\quad \psi_2=r^{-\eta/2}\frac{k}\omega(\delta g_x^y{}'-ik\delta g_r^y).
\end{equation}
The factors of $r$ included above allow us to decouple the equations of motion using the technique of ``master variables" \cite{kodama2004master}. In this approach, one finds a linear combination of the invariant variables $\psi_1$ and $\psi_2$ that will automatically decouple the equations. A brief discussion of how to find such variables is given in the Appendix (though no general algorithm is yet known). The answer is:
\begin{equation}
\phi_\pm=\psi_1+\lambda_\pm\psi_2
\end{equation}
where
\begin{equation}\label{X}
\lambda_\pm=\frac{1+\eta+k^2 \pm X}{k^2}\,,\quad X=\sqrt{(1+\eta)^2-2k^2(\zeta-1)}.
\end{equation}
In terms of these fields, we obtain two decoupled equations of motion:
\begin{equation}
0=\phi_\pm''+\phi_\pm\left[\frac{1-4\nu_\pm^2}{4r^2}+\omega^2\right], 
\end{equation}
with 
\begin{equation}\label{nu}
2\nu_\pm=\sqrt{5+2\eta+\eta^2+4k^2\pm 4X}.
\end{equation}
It is quite remarkable that the transverse perturbation equations can still be decoupled after the introduction of the mass term for the vector. As we will see below, we will not be so lucky in the longitudinal sector.
The decoupled equations are solved by Bessel functions
\begin{equation}
 \phi_\pm=c_1 \sqrt{r} J_{\nu_\pm }\left(\text{$\omega$}r \right)+c_2 \sqrt{r} Y_{\nu_\pm }\left(\text{$\omega$}r \right).
\end{equation}

We must choose only the ingoing modes at the horizon $(r\rightarrow\infty)$, since we know that matter falls into the black hole and does not come out. The solution satisfying these infalling boundary conditions is
\begin{equation}
\label{Hankel}
 \phi_\pm=\sqrt{r} H_{\nu }^{(1)}\left(\omega r\right).
\end{equation}

This is the near-horizon solution. We are ultimately interested in the retarded Green's function in the UV, $G^R(\omega,k)$. The imaginary part is the so-called spectral weight of a given operator, which counts the number of degrees of freedom that overlap with that operator, $\text{Im}G^R_{\mathcal{O}_A\mathcal{O}_A}(\omega,k)$. In general, to achieve this one must solve the equations of motion throughout the entire bulk, rather than just in the near horizon limit like we did. However, as shown in \cite{Donos:2012ra}, for the geometries we are considering it is possible to map the IR Green's function $\mathcal{G}^R(\omega,k)$ directly onto the UV one $G^R(\omega,k)$ through a matching procedure, provided we are only interested in the low frequency behaviour of the UV Green's function. The result is:
\begin{equation}
G^R(\omega,k)=d^0+\sum d^I \mathcal{G}_I^R(\omega,k),
\end{equation}
where the $d's$ are real constants and the sum runs over the fields involved. In our case this is just $\mathcal{G}^R_+$ and $\mathcal{G}^R_-$.

Returning to the IR solution, the Green's function is obtained by taking the limit of our near horizon solution as $r\rightarrow 0$. This corresponds to a solution far from the horizon:
\begin{equation}
\phi_{\pm}\propto\sqrt{r}(r^{-\nu_{\pm}}+\mathcal{G}^R_{\pm}(\omega,k)r^{\nu_{\pm}}),
\end{equation}
with
\begin{equation}
\mathcal{G}^R_{\pm}(\omega,k)\propto \omega^{2\nu_{\pm}}.
\end{equation}
We are now in a position to access the small $\omega$ dependence of the UV Green's function $G^R$:
\begin{equation}
G^R(\omega,k)=d^0+d^+ \mathcal{G}_+^R(\omega,k)+d^-\mathcal{G}_-^R(\omega,k).
\end{equation}
It is the imaginary part of this quantity that gives the low energy limit of the spectral weight, which is the quantity we wish to compute. Since the constants in are all real, and because the $\nu_-$ exponent dominates over the $\nu_+$ exponent at low energies (cf. (\ref{nu})), we obtain
\begin{equation}
\text{Im}G^R\propto\omega^{2\nu_-}. 
\end{equation} 
We note at this point that $\nu_\pm$ are both real in the allowed parameter space \eqref{ParSpace}.

Thus we find the following low energy spectral weight:
\begin{equation}
\sigma(k)=\lim_{\omega\to 0}\frac{\text{Im}G^R_{JJ}(\omega,k)}{\omega}=\left\{
\begin{array}{ll}
\infty \qquad & k<{k_\star}\\ ~ 0\qquad & k>{k_\star}
\end{array}
\right.
\end{equation}
where
\begin{equation}
{k_\star}^2=\frac{1}{4} \left(2 \sqrt{4 \zeta ^2+2 \zeta  \eta  (\eta +2)+2 \eta  (\eta +2)}-4 \zeta -\eta  (\eta +2)\right).
\end{equation}
This is a similar conclusion to the one reached in \cite{anantua2012pauli}. At zero temperature, this system enjoys low energy spectral weight over a finite range of momenta. This result runs contrary to weak coupling intuition, where we do not expect low energy degrees of freedom at finite momentum when the charges manifestly form a condensate. At strong coupling, gapless degrees of freedom can survive thanks to the scaling geometry in the IR. 

\begin{figure}
\centering
\includegraphics[width=.5\textwidth]{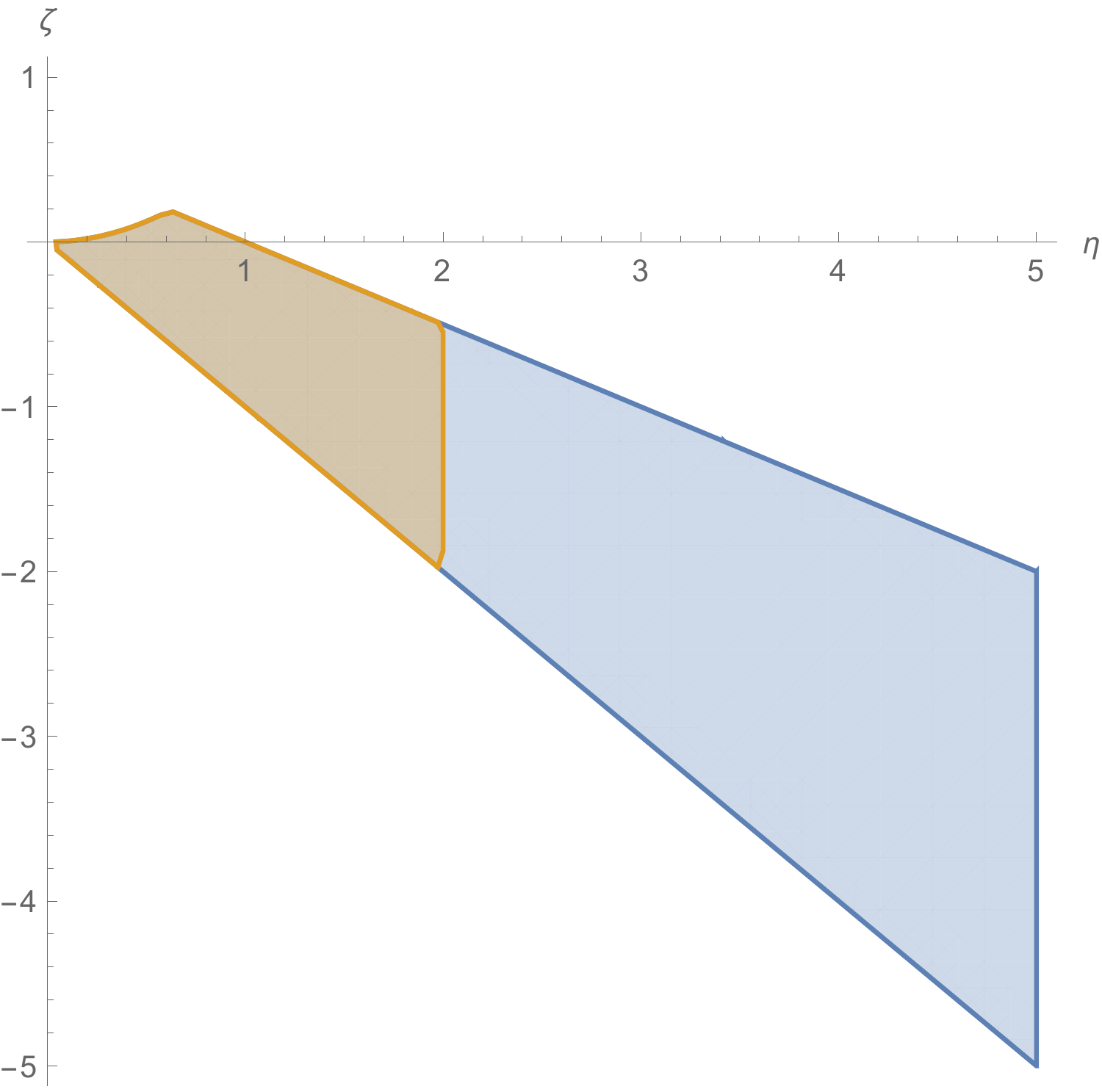}
\caption{Parameter space $(\eta,\zeta)$ where the low energy transverse spectral weight diverges for $k\le k_\star$ (brown region). The union of the two colored regions shows the full parameter space.}
\label{fig:ParSpaceTransverse}
\end{figure}

Notice, however, that $k_\star$ vanishes for the special value $\eta=2$ (independent of the value of $\zeta$). For $\eta>2$, there is never any low-energy spectral weight. It is worth recalling that without a condensate, $k_\star$ also vanishes for $\eta=2$ \cite{anantua2012pauli}. The full parameter space where the transverse spectral weight diverges at low frequency is depicted in Figure \ref{fig:ParSpaceTransverse}. We can also see that $\zeta>-2$ in order for the low energy transverse spectral weight to diverge.

There is an alternate method of obtaining the $\omega$ dependence of the Green's function without the necessity of decoupling the equations of motion. We will demonstrate this technique here for the transverse channel, and will then take advantage of it in the computationally complicated longitudinal channel. 

In this method, we endow all perturbations with a scaling behavior:
\begin{align}
\label{TransverseScalingAnsatz}
\begin{split}
&\delta A_y(r)=a_0r^{a_1},\quad \delta g_{ty}(r)=t_0r^{t_1}, \\
&\delta g_{xy}(r)=x_0r^{x_1},\quad\delta g_{ry}(r)=0.
\end{split}
\end{align}
Upon substituting this Ansatz into the four perturbed equations of motion, we notice that by fixing $t_1$ and $x_1$ in terms of $a_1$,\footnote{The specific relation between the exponents can be guessed simply by scaling analysis.} the equations take the schematic form $A(a_0,t_0,x_0)+B(a_0,t_0,x_0)\omega^2 r^2$. Namely, all the $r$ and $\omega$ dependence is relegated in the second term. This means we can set $\omega=0$ and simply scale $r$ to reinstate it. This is in agreement to the exact solution \eqref{Hankel} we found using the master variables. By manipulating the equations, we can solve for two of $(a_0,x_0,t_0)$ in terms of a third. Finally, the remaining equation can be solved to get an expression for $a_1$ in terms of $\eta$ and $\zeta$, yielding\footnote{Note that this method also finds spurious, pure gauge modes. They typically have no $k$ dependence and correspond for instance to rescalings of $t$, $x$ or $y$.}
\begin{align}
&a_1=\frac{1}{2} \left(2 \zeta +\eta+1  \pm \sqrt{5+\eta ^2+2 \eta +4 k^2 \pm 4 X}\right),
\end{align}
where $X$ is defined above in (\ref{X}). Note that the radical is the same as that obtained in (\ref{nu}), but here we avoided using the full machinery of gauge invariant variables and decoupling the equations of motion.

\subsection{Longitudinal channel}

In the longitudinal sector, we could not succeed in decoupling the field equations in terms of gauge invariant variables. We could not even express the equations solely in terms of gauge invariant variables. This is likely due to the fact that a $\delta a_t$ perturbation is turned on, which directly affects the right-hand side of the $t$ component of Maxwell equations via the mass term for $A^\mu$. Nonetheless, we can still work out the scaling dimensions of the IR operators according to the method outlined at the end of the previous section. 

In more details, the constraints coming from Einstein equations can be used to solve for $\delta a_x$ and $\delta g_{tx}$. Upon replacing and substituting for the longitudinal fluctuations an Ansatz similar to \eqref{TransverseScalingAnsatz}, the remaining equations all take the form $A(a_0,t_0,x_0)+B(a_0,t_0,x_0)\omega^2 r^2$, as in the transverse sector. Applying similar manipulations to this set of equations, after setting $\omega=0$, we find the following IR scaling dimensions:

\begin{equation}
\begin{split}
&\nu^{0}=\frac12\sqrt{(1+\eta)^2+4k^2}\\
&\nu^\pm=\frac{\sqrt{24 \zeta ^2-16 \zeta ^3-16 \zeta ^2 \eta -14 \zeta  \eta ^2+8 \zeta  \eta -10 \zeta +\eta ^4-2 \eta ^3+5 \eta ^2-4 k^2 \left(2 \zeta -\eta ^2\right)\pm4\sqrt X}}{2\sqrt{\eta^2-2\zeta}}\\
&X=(2 \zeta +\eta -1)^2 \left(2 \zeta ^2+\zeta  \eta -2 \zeta +\eta ^2\right)^2+4 (\zeta -1) k^2 \left(2 \zeta -\eta ^2\right) \left(2 \zeta ^2+2 \zeta  \eta -\zeta +\eta ^2\right)
\end{split}
\end{equation}
We can check that the dominant exponent at low frequencies is always $\nu^-$. Setting $W_0=0$ (or equivalently $\zeta=-\eta$), we recover the result reported in \cite{anantua2012pauli}. In this limit,  \cite{anantua2012pauli} found no low energy longitudinal spectral weight, that is $2\nu^--1$ is always positive. 

\begin{figure}
\centering
\includegraphics[width=.45\textwidth]{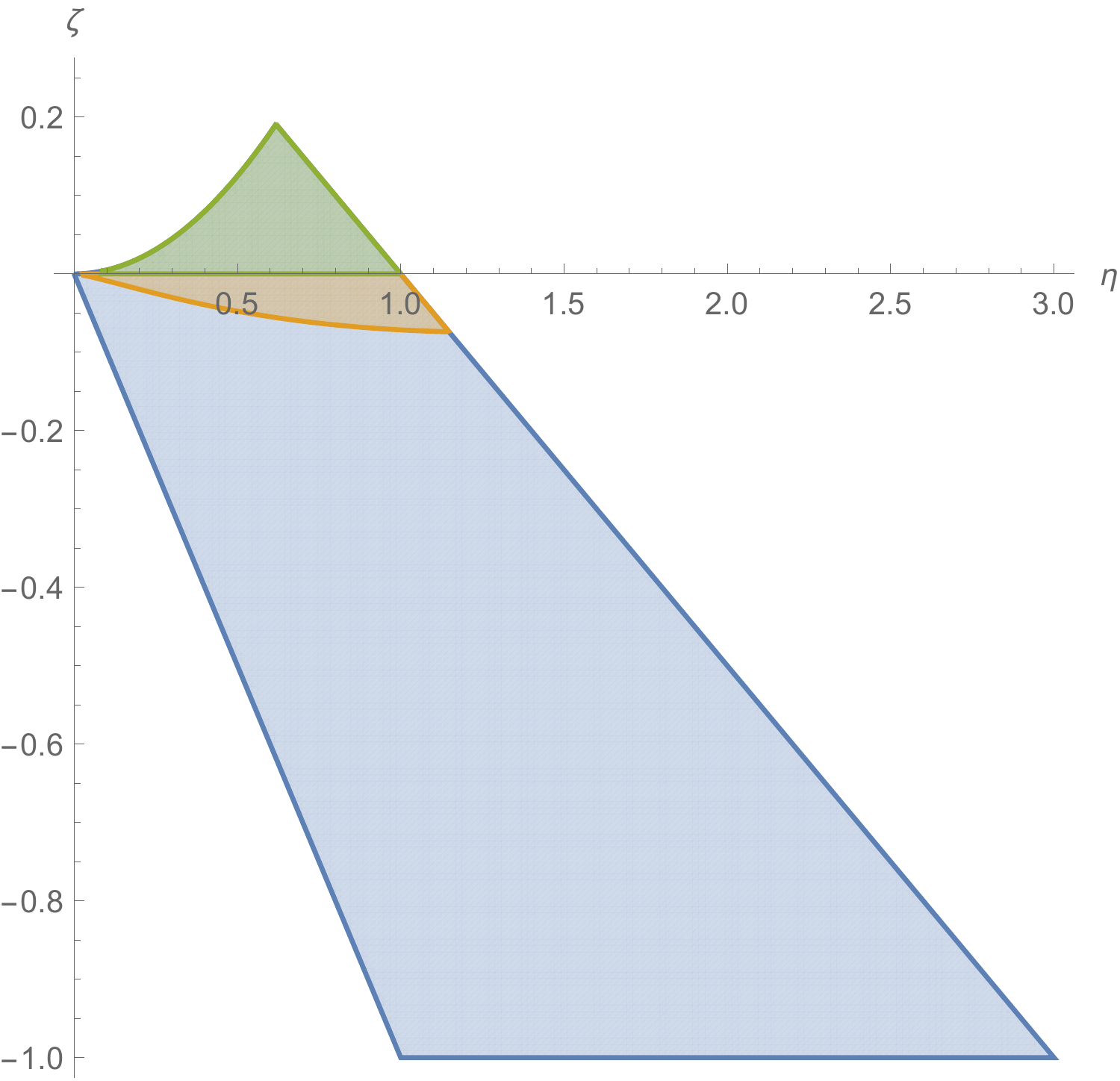}
\caption{In the green region, the mode $\nu^-(k)$ is complex for a range of wavevectors $k_-<k<k_+$, while it is real in the brown and blue regions for all values of $k$. In the brown region, $2\nu^--1<0$ for a range of wavevectors $k^\star_-<k<k^\star_+$ while in the blue region $2\nu^--1>0$, for any real $k$. The brown region of this plot is entirely contained within the brown region of Figure \ref{fig:ParSpaceTransverse}.}
\label{fig:parspacelong}
\end{figure}

There are several interesting differences when $W_0\neq0$. First, while $\nu^0$ and $\nu^+$ are always real in the allowed parameter space \eqref{ParSpace}, there is a region where $\nu^-$ can be complex for a range of $k$:
\begin{equation}
\left[0<\eta \leq \frac{1}{2} \left(\sqrt{5}-1\right)\; \textrm{and}\;  0<\zeta <\frac{\eta ^2}{2}\right] \textrm{or} \left[\frac{1}{2} \left(\sqrt{5}-1\right)<\eta <1\;\textrm{and}\; 0<\zeta <\frac{1-\eta }{2}\right]
\end{equation}
and $k_-<k<k_+$, where $(\nu^-(k_\pm))^2=0$. Effectively, this restricts $\zeta<0$, see Figure \ref{fig:parspacelong}. This suggests the homogeneous superfluid phase is unstable and that the endpoint of this instability is a spatially modulated, superfluid phase. To confirm its presence, we would need to construct the superfluid geometry at finite temperature and look for normalizable modes at non-zero $k$, along the lines of \cite{Donos:2011bh}. Spatially modulated holographic superfluids have been constructed before in various other setups \cite{Donos:2011ff,Donos:2012gg,Donos:2013woa,Krikun:2013iha,Erdmenger:2013zaa,Krikun:2015tga,Cremonini:2016rbd}. In this region of parameter space, there is always a corresponding Fermi surface in the transverse sector.

Also in Figure \ref{fig:parspacelong}, we show in the brown region the parameter space $(\eta,\zeta)$ where $2\nu^--1$ is negative for $k^\star_-<k<k^\star_+$ (where $2\nu^-(k^\star_\pm)-1=0$). When $k<k_-^\star$ or $k>k^\star_+$, then $2\nu^--1>0$. This suggests we do not have a Fermi surface and corresponding Fermi sea for $k\leq k_F$, but rather a Fermi shell for $k^\star_-<k<k^\star_+$. This shell may result from two nested Fermi surfaces, one of `particles' and one of `holes'. When the longitudinal sector exhibits this shell, there is always a corresponding Fermi surface in the transverse sector. Let us note here that Fermi shells have also appeared in top-down truncations, both in $\mathcal N=4$ SYM \cite{DeWolfe:2012uv} and ABJM theory \cite{DeWolfe:2014ifa}. There, the situation is clearer as the field content of the dual field theory is known and there are indeed two species of fermions, each with their own $k_F$ of opposite sign. The superposition of both Fermi surfaces leads to a Fermi shell. Our bottom-up setup unfortunately does not allow such a precise identification of the degrees of freedom.

\section{Discussion}

We find that holographic superfluids with a semi-locally critical IR can exhibit low energy spectral weight over a finite range of momenta in the transverse correlators, which is a signature of a (smeared) Fermi surface. This result is surprising, as the bulk charge density that is thought to be responsible for this spectral weight manifestly forms a condensate in our model. The formation of the condensate is not accompanied by a gapping out of the low energy degrees of freedom.

In the longitudinal sector, we find that there is no low-energy spectral weight in a large part of the parameter space. Interestingly, we do find a region where there exists a Fermi `shell', which suggests the presence of two Fermi surfaces, one of `holes' and one of `electrons'. In yet another region of the parameter space, we find that the scaling dimension of the least irrelevant IR operator becomes complex for non-zero values of the wavevector, indicating an instability, the endpoint of which is likely to be a spatially modulated superfluid phase. Obviously, it would be very interesting to construct these new ground states. Both of these features are always accompanied by non-zero spectral weight in the transverse sector, although the converse is not true.

While we have focussed on the case where all of the bulk charge sits outside the horizon, superfluid $\eta$ geometries with charged horizons can also be constructed \cite{gouteraux2012quantum}. Their spectral weight is computed by same scaling arguments as developed here, and would be characterized by the same $\nu(k)$ exponents as in \cite{anantua2012pauli}. The reason for this being that the effects of the mass term for the vector are subleading in the IR and so do not affect the IR Green's function at low frequency. Thus the same distribution of spectral weight as in \cite{anantua2012pauli} would be found: transverse spectral weight provided $0<\eta<2$, but no longitudinal spectral weight.

Our results raise some interesting questions about spectral weight in holographic theories, such as: a) To what extent do charge distribution properties in the gravity theory represent those of the field theory? More precisely, it would be desirable to understand more clearly the relation between bulk charge carried by the horizon or a condensate, and the normal and superfluid charge densities on the boundary. Preliminary work in this direction appeared in \cite{Sonner:2010yx}. b) How should one interpret spectral weight at finite wavevector when all the charge in the bulk is manifestly carried by a condensate? and c) Why does the presence of low energy spectral weight seem to depend more upon the background geometry than matter content?

To work toward answering these questions, it may be important to recall that the spectral weight found in the holographic superfluid case vanishes for specific values of the background parameters of the theory, namely $\eta=2$. Interestingly, a recent work has uncovered an example of a theory such that, after being dimensionally reduced from a higher dimensional Supergravity theory, the Fermi surface present in fermionic correlators vanishes when the parameters in the holographic action take their Supergravity values \cite{dewolfe2016gapped}. Moreover these top-down constructions have the advantage over bottom-up ones in that they have a well-identified field theory dual. 

Our results also connect with work on the presence of holographic Friedel oscillations in the spatially resolved static susceptibility $\chi(k)$, either for Reissner - Nordstrom AdS \cite{Blake:2014lva} or the $U(1)^4$ black hole of \cite{Cvetic:1999xp} with three of the charges equal and the fourth zero. In the former case, exponentially damped Friedel oscillations were uncovered at long distances, apparently at the wavevector where $Re[\nu^-(k)]=0$ (with the expressions appropriate for an AdS$\times R^2$ IR geometry). \cite{Henriksson:2016gfm} found that, on the contrary, these oscillations do not survive at long distances in the $U(1)^4$ black hole, which happens to have an $\eta=1$ IR. This is consistent with the results in \cite{anantua2012pauli} that there is no low energy longitudinal spectral weight. It would be worthwhile to compute the static susceptibility in our superfluid geometries and work out whether Friedel oscillations are present in the corner of parameter space where we do find longitudinal spectral weight.

\acknowledgments
We would like to thank Chris Rosen for very interesting discussions and for sharing with us the results of \cite{Henriksson:2016gfm} prior to their publication. We are also grateful to Sean Hartnoll for discussions and comments on a draft. BG is partially supported by the Marie Curie International Outgoing Fellowship nr 624054 within the 7th European Community Framework Programme FP7/2007-2013. The work of BG was partially performed at the Aspen Center for Physics, which is supported by National Science Foundation grant PHY-1066293. BG would also like to acknowledge many interesting discussions there with Jan Zaanen on related topics. VM is supported by the DARE Stanford Graduate Fellowship Program. 

\appendix

\section{Field Ansatz and equations of motion \label{AppEoms}}
 We begin with the general background fields ansatz:
\begin{equation}
\label{eq:EMDFieldAnsatz}
ds^2=B(r)dr^2+C(r)\left(d\vec{x}^2\right)-D(r)dt^2\,,\quad A_t(r)=A(r)\,,\quad \phi=\phi(r).
\end{equation}
The background equations following from this ansatz are the Maxwell equation
\begin{equation}
\label{eq:EMDMaxt}
0=\frac1C\sqrt{\frac{D}{B}}\left(\frac{Z C}{\sqrt{BD}}A'\right)'+W A,
\end{equation}
the scalar equation
\begin{equation}
\label{eq:EMDScal}
0=\sqrt{\frac{B}{D}}C^{-1}\left(C\sqrt{\frac{D}{B}}\phi'\right)'+Z_{,\phi}\frac{(A')^2}{2D}-B V_{,\phi}+\frac{B A^2}{2D}W_{,\phi},
\end{equation}
and the Einstein equations
\begin{equation}
\label{eq:EMDEE1}
0=\sqrt{\frac{B}{D}}C^{-1}\left(\frac{C}{\sqrt{BD}}D'\right)'+BV-\frac{1}{2}\frac{Z(A')^2}{D}-\frac{B}{D}W A^2,
\end{equation}
\begin{equation}
\label{eq:EMDEE2}
0=\sqrt{\frac{BD}{C}}\left(\frac{C'}{\sqrt{BCD}}\right)'+\frac12(\phi')^2+\frac{B}{2D}W A^2,
\end{equation}
\begin{equation}
\label{eq:EMDEE3}
0=(\phi')^2-\frac{C'}{C}\left(2\frac{D'}D+\frac{C'}C\right)-\frac{Z(A')^2}{D}+\frac{B}{D}W A^2-2BV.
\end{equation}
There is a background conserved quantity
\begin{equation}
\label{eq:EMDConservedQtyBackground}
0=\left[\frac{C}{\sqrt{BD}}\left(ZAA'-C\left(\frac{D}C\right)'\right)\right]'
\end{equation}
which gives the Smarr law when evaluated at the boundary and at the horizon.

\section{Lie derivatives}
Here we calculate the Lie derivative of the metric and gauge field along a vector field $\xi$. For the transverse channel, we choose our vector field so that only the y-component is nonzero: $\xi=e^{i(kx-\omega t)}\xi^y(r)\partial_y$. This ensures that it too is odd under $y\rightarrow-y$. For the metric perturbations, the Lie derivative is given by 
\begin{equation}
[\mathcal{L}_\xi g]_{\mu\nu}=\xi^\lambda\partial_\lambda g_{\mu\nu}+ g_{\lambda\nu}\partial_\mu\xi^\lambda+ g_{\mu\lambda}\partial_{\nu}\xi^\lambda.
\end{equation}
Our metric was found to be 
\begin{equation}
ds^2=r^{-\eta}\left(\frac{-dt^2+L^2 dr^2}{r^2}+dx^2+dy^2\right),
\end{equation}
so we can directly calculate
\begin{equation}
\begin{split}
&[\mathcal{L}_\xi g]_{yt}=g_{yy}\partial_t\xi^y=-i\omega r^{-\eta}\xi^y\\
&[\mathcal{L}_\xi g]_{yr}=g_{yy}\partial_r\xi^y=r^{-\eta}\xi^{'y}\\
&[\mathcal{L}_\xi g]_{yx}=g_{yy}\partial_x\xi^y=ik r^{-\eta}\xi^y.
\end{split}
\end{equation}
To get rid of the factors $r^{-\eta}$, it is common to work in one raised index, so that
\begin{equation}\label{Lie deriv}
\begin{split}
&[\mathcal{L}_\xi g]^{y}_t=-i\omega \xi^y, \quad
[\mathcal{L}_\xi g]^{y}_r=\xi^{'y}, \quad
[\mathcal{L}_\xi g]^y_{x}=ik \xi^y.
\end{split}
\end{equation}
The Lie derivative of the gauge field is
\begin{equation}
[\mathcal{L}_\xi a]_{y}=\xi^\lambda\partial_\lambda a_{y}+ a_{\lambda}\partial_y\xi^\lambda=0.
\end{equation}

\section{Finding master variables}
By inspecting (\ref{Lie deriv}), we can identify a good choice for gauge invariant variables, up to factors of $r$:
\begin{equation}
\psi_1=r^{N_1}(\delta g_t^y{}'+i\omega \delta g_r^y)\,,\quad \psi_2=r^{N_2}\frac{k}\omega(\delta g_x^y{}'-ik\delta g_r^y).
\end{equation}
The $k/\omega$ factor is just for ascetics later. We can now write the perturbation equations of motion in terms of these variables:
\begin{multline}\label{PertEq1}
0=\psi_2(r) \left(-2\zeta-2\eta-k^2\right) r^{N_1-N_2-4}+\frac{\psi_1(r) \left(-3 \eta -2k^2+N_1 (\eta
	+N_1-3)+r^2 \omega ^2\right)}{r^2}\\-\frac{(\eta +2 N_1-4) \psi_1'(r)}{r}+\psi_1''(r)
\end{multline}
\begin{align}\label{PertEq2}
0=k^2 \psi _1(r) r^{N_2-N_1}+\frac{\psi_2(r) \left(\eta +N_2^2+\eta N_2+N_2+r^2 \omega ^2\right)}{r^2}-\frac{(\eta +2
	N_2) \psi_2'(r)}{r}+\psi_2''(r).
\end{align}

The master variables will be some linear combination of the $\psi$'s. A general combination would be 
\begin{equation}
\phi_{\pm}=f_{1\pm}(r)\psi_1(r)+f_{2\pm}(r)\psi_2(r)+f_{3\pm}(r)\psi_1'(r)+f_{4\pm}(r)\psi_2'(r),
\end{equation}
but we would like a simpler choice. A good practice is to only include two of the terms that occur in both coupled equations. In the present case, this rules out $\psi_1'$ and $\psi_2'$, so we have 
\begin{equation}
\phi_{\pm}=f_{1\pm}(r)\psi_1(r)+f_{2\pm}(r)\psi_2(r).
\end{equation}
Now it remains to solve for the functions $f_{1\pm}(r)$ and $f_{2\pm}(r)$ that satisfy $\phi_{\pm}''\propto \phi_{\pm}$. We will see how this condition is also used to fix the exponents $N_1$ and $N_2$. To that end, let's take a look at $\phi_{\pm}''$:
\begin{equation}
\phi_{\pm}''=f_{1\pm}''(r)\psi_1(r)+2f_{1\pm}'(r)\psi_1'(r)+f_{1\pm}(r)\psi_1''(r)+f_{2\pm}''(r)\psi_2(r)+2f_{2\pm}'(r)\psi_2'(r)+f_{2\pm}(r)\psi_2''(r)
\end{equation}
We can then use (\ref{PertEq1}) and (\ref{PertEq2}) two get rid of the second derivatives in favor of lower order terms. Schematically, then, $\phi_{\pm}''$ becomes
\begin{equation}
\phi_{\pm}''=g_{1\pm}(r)\psi_1(r)+g_{2\pm}(r)\psi_1'(r)+g_{3\pm}(r)\psi_2(r)+g_{4\pm}(r)\psi_2'(r).
\end{equation}
To satisfy the condition $\phi_{\pm}''\propto \phi_{\pm}$, we see immediately that the coefficient functions of $\psi_1'(r)$ and $\psi_2'(r)$ should be set to zero. This yields two differential equations, one in $f_{1\pm}(r)$ and the other in $f_{2\pm}(r)$, and the exponents $N_1$ and $N_2$ are chosen so that these two functions are constants. From there it is easy to determine what these constants need to be in order to achieve $\phi_{\pm}''\propto \phi_{\pm}$. As we noted earlier, it is not always possible to find simple solutions to these differential equations, and as yet there are no criteria to determine from the beginning which coupled equations can be decoupled using master variables and which cannot. 

\bibliographystyle{jhep}
\bibliography{SpecBib}

\providecommand{\href}[2]{#2}\begingroup\raggedright\begin{thebibliography}{10}

\bibitem{varma1989phenomenology}
C.~Varma, P.~B. Littlewood, S.~Schmitt-Rink, E.~Abrahams and A.~Ruckenstein,
  \emph{Phenomenology of the normal state of cu-o high-temperature
  superconductors}, {\emph{Physical Review Letters} {\bf 63} (1989) 1996}.

\bibitem{schofield1999non}
A.~J. Schofield, \emph{Non-fermi liquids}, {\emph{Contemporary Physics} {\bf
  40} (1999) 95--115}.

\bibitem{hartnoll2009lectures}
S.~A. Hartnoll, \emph{{Lectures on holographic methods for condensed matter
  physics}},
  \href{http://dx.doi.org/10.1088/0264-9381/26/22/224002}{\emph{Class. Quant.
  Grav.} {\bf 26} (2009) 224002}, [\href{https://arxiv.org/abs/0903.3246}{{\tt
  0903.3246}}].

\bibitem{Charmousis:2010zz}
C.~Charmousis, B.~Gout\'eraux, B.~S. Kim, E.~Kiritsis and R.~Meyer,
  \emph{{Effective Holographic Theories for low-temperature condensed matter
  systems}}, \href{http://dx.doi.org/10.1007/JHEP11(2010)151}{\emph{JHEP} {\bf
  11} (2010) 151}, [\href{https://arxiv.org/abs/1005.4690}{{\tt 1005.4690}}].

\bibitem{Gouteraux:2011ce}
B.~Gout\'eraux and E.~Kiritsis, \emph{{Generalized Holographic Quantum
  Criticality at Finite Density}},
  \href{http://dx.doi.org/10.1007/JHEP12(2011)036}{\emph{JHEP} {\bf 12} (2011)
  036}, [\href{https://arxiv.org/abs/1107.2116}{{\tt 1107.2116}}].

\bibitem{huijse2012hidden}
L.~Huijse, S.~Sachdev and B.~Swingle, \emph{{Hidden Fermi surfaces in
  compressible states of gauge-gravity duality}},
  \href{http://dx.doi.org/10.1103/PhysRevB.85.035121}{\emph{Phys. Rev.} {\bf
  B85} (2012) 035121}, [\href{https://arxiv.org/abs/1112.0573}{{\tt
  1112.0573}}].

\bibitem{hartnoll2012spectral}
S.~A. Hartnoll and E.~Shaghoulian, \emph{{Spectral weight in holographic
  scaling geometries}},
  \href{http://dx.doi.org/10.1007/JHEP07(2012)078}{\emph{JHEP} {\bf 07} (2012)
  078}, [\href{https://arxiv.org/abs/1203.4236}{{\tt 1203.4236}}].

\bibitem{anantua2012pauli}
R.~J. Anantua, S.~A. Hartnoll, V.~L. Martin and D.~M. Ramirez, \emph{{The Pauli
  exclusion principle at strong coupling: Holographic matter and momentum
  space}}, \href{http://dx.doi.org/10.1007/JHEP03(2013)104}{\emph{JHEP} {\bf
  03} (2013) 104}, [\href{https://arxiv.org/abs/1210.1590}{{\tt 1210.1590}}].

\bibitem{Iqbal:2011in}
N.~Iqbal, H.~Liu and M.~Mezei, \emph{{Semi-local quantum liquids}},
  \href{http://dx.doi.org/10.1007/JHEP04(2012)086}{\emph{JHEP} {\bf 04} (2012)
  086}, [\href{https://arxiv.org/abs/1105.4621}{{\tt 1105.4621}}].

\bibitem{gubser2008breaking}
S.~S. Gubser, \emph{{Breaking an Abelian gauge symmetry near a black hole
  horizon}}, \href{http://dx.doi.org/10.1103/PhysRevD.78.065034}{\emph{Phys.
  Rev.} {\bf D78} (2008) 065034}, [\href{https://arxiv.org/abs/0801.2977}{{\tt
  0801.2977}}].

\bibitem{hartnoll2008building}
S.~A. Hartnoll, C.~P. Herzog and G.~T. Horowitz, \emph{{Building a Holographic
  Superconductor}},
  \href{http://dx.doi.org/10.1103/PhysRevLett.101.031601}{\emph{Phys. Rev.
  Lett.} {\bf 101} (2008) 031601}, [\href{https://arxiv.org/abs/0803.3295}{{\tt
  0803.3295}}].

\bibitem{hartnoll2008holographic}
S.~A. Hartnoll, C.~P. Herzog and G.~T. Horowitz, \emph{{Holographic
  Superconductors}},
  \href{http://dx.doi.org/10.1088/1126-6708/2008/12/015}{\emph{JHEP} {\bf 12}
  (2008) 015}, [\href{https://arxiv.org/abs/0810.1563}{{\tt 0810.1563}}].

\bibitem{hartnoll2011horizons}
S.~A. Hartnoll, \emph{{Horizons, holography and condensed matter}},
  \href{https://arxiv.org/abs/1106.4324}{{\tt 1106.4324}}.

\bibitem{Gubser:2009cg}
S.~S. Gubser and A.~Nellore, \emph{{Ground states of holographic
  superconductors}},
  \href{http://dx.doi.org/10.1103/PhysRevD.80.105007}{\emph{Phys. Rev.} {\bf
  D80} (2009) 105007}, [\href{https://arxiv.org/abs/0908.1972}{{\tt
  0908.1972}}].

\bibitem{Horowitz:2009ij}
G.~T. Horowitz and M.~M. Roberts, \emph{{Zero Temperature Limit of Holographic
  Superconductors}},
  \href{http://dx.doi.org/10.1088/1126-6708/2009/11/015}{\emph{JHEP} {\bf 11}
  (2009) 015}, [\href{https://arxiv.org/abs/0908.3677}{{\tt 0908.3677}}].

\bibitem{gouteraux2012quantum}
B.~Gout\'eraux and E.~Kiritsis, \emph{{Quantum critical lines in holographic
  phases with (un)broken symmetry}},
  \href{http://dx.doi.org/10.1007/JHEP04(2013)053}{\emph{JHEP} {\bf 04} (2013)
  053}, [\href{https://arxiv.org/abs/1212.2625}{{\tt 1212.2625}}].

\bibitem{Gouteraux:2013oca}
B.~Goutéraux, \emph{{Universal scaling properties of extremal cohesive
  holographic phases}},
  \href{http://dx.doi.org/10.1007/JHEP01(2014)080}{\emph{JHEP} {\bf 01} (2014)
  080}, [\href{https://arxiv.org/abs/1308.2084}{{\tt 1308.2084}}].

\bibitem{Iizuka:2012pn}
N.~Iizuka, S.~Kachru, N.~Kundu, P.~Narayan, N.~Sircar, S.~P. Trivedi et~al.,
  \emph{{Extremal Horizons with Reduced Symmetry: Hyperscaling Violation,
  Stripes, and a Classification for the Homogeneous Case}},
  \href{http://dx.doi.org/10.1007/JHEP03(2013)126}{\emph{JHEP} {\bf 03} (2013)
  126}, [\href{https://arxiv.org/abs/1212.1948}{{\tt 1212.1948}}].

\bibitem{Gath:2012pg}
J.~Gath, J.~Hartong, R.~Monteiro and N.~A. Obers, \emph{{Holographic Models for
  Theories with Hyperscaling Violation}},
  \href{http://dx.doi.org/10.1007/JHEP04(2013)159}{\emph{JHEP} {\bf 04} (2013)
  159}, [\href{https://arxiv.org/abs/1212.3263}{{\tt 1212.3263}}].

\bibitem{Gubser:2009qm}
S.~S. Gubser, C.~P. Herzog, S.~S. Pufu and T.~Tesileanu, \emph{{Superconductors
  from Superstrings}},
  \href{http://dx.doi.org/10.1103/PhysRevLett.103.141601}{\emph{Phys. Rev.
  Lett.} {\bf 103} (2009) 141601}, [\href{https://arxiv.org/abs/0907.3510}{{\tt
  0907.3510}}].

\bibitem{Gauntlett:2009dn}
J.~P. Gauntlett, J.~Sonner and T.~Wiseman, \emph{{Holographic superconductivity
  in M-Theory}},
  \href{http://dx.doi.org/10.1103/PhysRevLett.103.151601}{\emph{Phys. Rev.
  Lett.} {\bf 103} (2009) 151601}, [\href{https://arxiv.org/abs/0907.3796}{{\tt
  0907.3796}}].

\bibitem{Gauntlett:2009bh}
J.~P. Gauntlett, J.~Sonner and T.~Wiseman, \emph{{Quantum Criticality and
  Holographic Superconductors in M-theory}},
  \href{http://dx.doi.org/10.1007/JHEP02(2010)060}{\emph{JHEP} {\bf 02} (2010)
  060}, [\href{https://arxiv.org/abs/0912.0512}{{\tt 0912.0512}}].

\bibitem{Bobev:2011rv}
N.~Bobev, A.~Kundu, K.~Pilch and N.~P. Warner, \emph{{Minimal Holographic
  Superconductors from Maximal Supergravity}},
  \href{http://dx.doi.org/10.1007/JHEP03(2012)064}{\emph{JHEP} {\bf 03} (2012)
  064}, [\href{https://arxiv.org/abs/1110.3454}{{\tt 1110.3454}}].

\bibitem{DeWolfe:2015kma}
O.~DeWolfe, S.~S. Gubser, O.~Henriksson and C.~Rosen, \emph{{Fermionic Response
  in Finite-Density ABJM Theory with Broken Symmetry}},
  \href{http://dx.doi.org/10.1103/PhysRevD.93.026001}{\emph{Phys. Rev.} {\bf
  D93} (2016) 026001}, [\href{https://arxiv.org/abs/1509.00518}{{\tt
  1509.00518}}].

\bibitem{dewolfe2016gapped}
O.~DeWolfe, S.~S. Gubser, O.~Henriksson and C.~Rosen, \emph{{Gapped Fermions in
  Top-down Holographic Superconductors}},
  \href{https://arxiv.org/abs/1609.07186}{{\tt 1609.07186}}.

\bibitem{Gouteraux:2014hca}
B.~Gout\'eraux, \emph{{Charge transport in holography with momentum
  dissipation}}, \href{http://dx.doi.org/10.1007/JHEP04(2014)181}{\emph{JHEP}
  {\bf 04} (2014) 181}, [\href{https://arxiv.org/abs/1401.5436}{{\tt
  1401.5436}}].

\bibitem{Karch:2014mba}
A.~Karch, \emph{{Conductivities for Hyperscaling Violating Geometries}},
  \href{http://dx.doi.org/10.1007/JHEP06(2014)140}{\emph{JHEP} {\bf 06} (2014)
  140}, [\href{https://arxiv.org/abs/1405.2926}{{\tt 1405.2926}}].

\bibitem{PhysRevB.46.2655}
X.-G. Wen, \emph{Scaling theory of conserved current and universal amplitudes
  at anisotropic critical points},
  \href{http://dx.doi.org/10.1103/PhysRevB.46.2655}{\emph{Phys. Rev. B} {\bf
  46} (Aug, 1992) 2655--2662}.

\bibitem{dong2012aspects}
X.~Dong, S.~Harrison, S.~Kachru, G.~Torroba and H.~Wang, \emph{Aspects of
  holography for theories with hyperscaling violation}, {\emph{Journal of High
  Energy Physics} {\bf 2012} (2012) 1--33}.

\bibitem{kodama2004master}
H.~Kodama and A.~Ishibashi, \emph{{Master equations for perturbations of
  generalized static black holes with charge in higher dimensions}},
  \href{http://dx.doi.org/10.1143/PTP.111.29}{\emph{Prog. Theor. Phys.} {\bf
  111} (2004) 29--73}, [\href{https://arxiv.org/abs/hep-th/0308128}{{\tt
  hep-th/0308128}}].

\bibitem{Donos:2012ra}
A.~Donos and S.~A. Hartnoll, \emph{{Universal linear in temperature resistivity
  from black hole superradiance}},
  \href{http://dx.doi.org/10.1103/PhysRevD.86.124046}{\emph{Phys. Rev.} {\bf
  D86} (2012) 124046}, [\href{https://arxiv.org/abs/1208.4102}{{\tt
  1208.4102}}].

\bibitem{Donos:2011bh}
A.~Donos and J.~P. Gauntlett, \emph{{Holographic striped phases}},
  \href{http://dx.doi.org/10.1007/JHEP08(2011)140}{\emph{JHEP} {\bf 08} (2011)
  140}, [\href{https://arxiv.org/abs/1106.2004}{{\tt 1106.2004}}].

\bibitem{Donos:2011ff}
A.~Donos and J.~P. Gauntlett, \emph{{Holographic helical superconductors}},
  \href{http://dx.doi.org/10.1007/JHEP12(2011)091}{\emph{JHEP} {\bf 12} (2011)
  091}, [\href{https://arxiv.org/abs/1109.3866}{{\tt 1109.3866}}].

\bibitem{Donos:2012gg}
A.~Donos and J.~P. Gauntlett, \emph{{Helical superconducting black holes}},
  \href{http://dx.doi.org/10.1103/PhysRevLett.108.211601}{\emph{Phys. Rev.
  Lett.} {\bf 108} (2012) 211601}, [\href{https://arxiv.org/abs/1203.0533}{{\tt
  1203.0533}}].

\bibitem{Donos:2013woa}
A.~Donos, J.~P. Gauntlett and C.~Pantelidou, \emph{{Competing p-wave orders}},
  \href{http://dx.doi.org/10.1088/0264-9381/31/5/055007}{\emph{Class. Quant.
  Grav.} {\bf 31} (2014) 055007}, [\href{https://arxiv.org/abs/1310.5741}{{\tt
  1310.5741}}].

\bibitem{Krikun:2013iha}
A.~Krikun, \emph{{Charge density wave instability in holographic d-wave
  superconductor}},
  \href{http://dx.doi.org/10.1007/JHEP04(2014)135}{\emph{JHEP} {\bf 04} (2014)
  135}, [\href{https://arxiv.org/abs/1312.1588}{{\tt 1312.1588}}].

\bibitem{Erdmenger:2013zaa}
J.~Erdmenger, X.-H. Ge and D.-W. Pang, \emph{{Striped phases in the holographic
  insulator/superconductor transition}},
  \href{http://dx.doi.org/10.1007/JHEP11(2013)027}{\emph{JHEP} {\bf 11} (2013)
  027}, [\href{https://arxiv.org/abs/1307.4609}{{\tt 1307.4609}}].

\bibitem{Krikun:2015tga}
A.~Krikun, \emph{{Phases of holographic d-wave superconductor}},
  \href{http://dx.doi.org/10.1007/JHEP10(2015)123}{\emph{JHEP} {\bf 10} (2015)
  123}, [\href{https://arxiv.org/abs/1506.05379}{{\tt 1506.05379}}].

\bibitem{Cremonini:2016rbd}
S.~Cremonini, L.~Li and J.~Ren, \emph{{Holographic Pair and Charge Density
  Waves}},  \href{https://arxiv.org/abs/1612.04385}{{\tt 1612.04385}}.

\bibitem{DeWolfe:2012uv}
O.~DeWolfe, S.~S. Gubser and C.~Rosen, \emph{{Fermi surfaces in N=4
  Super-Yang-Mills theory}},
  \href{http://dx.doi.org/10.1103/PhysRevD.86.106002}{\emph{Phys. Rev.} {\bf
  D86} (2012) 106002}, [\href{https://arxiv.org/abs/1207.3352}{{\tt
  1207.3352}}].

\bibitem{DeWolfe:2014ifa}
O.~DeWolfe, O.~Henriksson and C.~Rosen, \emph{{Fermi surface behavior in the
  ABJM M2-brane theory}},
  \href{http://dx.doi.org/10.1103/PhysRevD.91.126017}{\emph{Phys. Rev.} {\bf
  D91} (2015) 126017}, [\href{https://arxiv.org/abs/1410.6986}{{\tt
  1410.6986}}].

\bibitem{Sonner:2010yx}
J.~Sonner and B.~Withers, \emph{{A gravity derivation of the Tisza-Landau Model
  in AdS/CFT}}, \href{http://dx.doi.org/10.1103/PhysRevD.82.026001}{\emph{Phys.
  Rev.} {\bf D82} (2010) 026001}, [\href{https://arxiv.org/abs/1004.2707}{{\tt
  1004.2707}}].

\bibitem{Blake:2014lva}
M.~Blake, A.~Donos and D.~Tong, \emph{{Holographic Charge Oscillations}},
  \href{http://dx.doi.org/10.1007/JHEP04(2015)019}{\emph{JHEP} {\bf 04} (2015)
  019}, [\href{https://arxiv.org/abs/1412.2003}{{\tt 1412.2003}}].

\bibitem{Cvetic:1999xp}
M.~Cvetic, M.~J. Duff, P.~Hoxha, J.~T. Liu, H.~Lu, J.~X. Lu et~al.,
  \emph{{Embedding AdS black holes in ten-dimensions and eleven-dimensions}},
  \href{http://dx.doi.org/10.1016/S0550-3213(99)00419-8}{\emph{Nucl. Phys.}
  {\bf B558} (1999) 96--126}, [\href{https://arxiv.org/abs/hep-th/9903214}{{\tt
  hep-th/9903214}}].

\bibitem{Henriksson:2016gfm}
O.~Henriksson and C.~Rosen, \emph{{"$1k_F$" Singularities and Finite Density
  ABJM Theory at Strong Coupling}},
  \href{https://arxiv.org/abs/1612.06823}{{\tt 1612.06823}}.

\end{thebibliography}\endgroup
\end{document}